\documentclass[%
reprint,                     
amsmath,amssymb,
aps,
]{revtex4-2}

\usepackage{graphicx}
\usepackage{dcolumn}
\usepackage{bm}

\usepackage{xcolor}
\usepackage{soul}

\usepackage{times}

\usepackage{comment}
\begin{document}
	
	\preprint{APS/123-QED}
	
	\title{ Negative Radiation Pressure Scheme for Simultaneous Suppression of Arduous Back-Action Evasion and Shot Noise in Gravitational Wave Detectors
 }
	
	\author{Souvik Agasti}
	\email{souvik.agasti@uhasselt.be}
	\author{Abhishek Shukla}%
	\author{Milos Nesladek}%
	
	\affiliation{Physics Department, Hasselt University 
	}%

	
    \begin{abstract}
		
    Aiming at application for gravitational wave (GW) detection, we propose a novel scheme how to obtain quantum back action evading measurements performed on an opto-mechanical cavity, by introducing a negative radiation pressure coupling between the cavity field and the end mirror. The scheme consists of introducing a double cavity with end mirrors interlocked by a pivot and moving in opposite directions. 
    The measurement is performed by sending a two-mode squeezed vacuum to both cavities and detecting the output through the heterodyne detection. Compared to the previously proposed hybrid negative mass spin-optomechanical system in \cite{study4roadmap}, we see that our scheme is capable to suppress back action noise by nearly two orders of magnitude more in the lower frequency region. Overall, the setup has been able to squeeze the output noise below the standard quantum limit, with more efficiency. In addition, the scheme has also proven to be beneficial for reducing thermal noise by a significant amount. We confirm our result by a numerical analysis and compared it with the previous proposal \cite{study4roadmap}.
		
	\end{abstract}
	
	\maketitle
	

	\section{Introduction}\label{Introduction}
	
   {The second-generation gravitational wave (GW) detectors (such as advanced LIGO \cite{advanced_LIGO}, advanced Virgo \cite{advanced_Virgo} and advanced GEO \cite{advanced_GEO}) and others aim at increasing the GW detection sensitivity and significantly reducing the detection noise. The noise is composed of mainly electronic, seismic, thermal, and quantum noise components. For example, thermal noise can be reduced using a cryogenic setup or low thermal noise coatings \cite{coating_thermal_noise}. At the current state of the art, the quantum noise is still the major concern of limitation \cite{Danilishin_GW_2019}. The arms of a GW detector are typically considered to be optomechanical systems. The main sources of quantum noise in optomechanical systems are the shot noise, which is dependent on the number of incident photons injected into the system $\delta N \propto \sqrt{N}$, and the quantum back-action (QBA) noise which is generated due to the optomechanical interaction of the cavity field with the end mirror. 
   Further progress in the quantum noise suppression in GW interferometers has been achieved by using homodyne detection.
   In this case, the essence of quantum noise and its spectral dependence is mathematically encapsulated in the quantum noise spectral density, described in terms of the expectation value of quadrature field operators as $ S_a (\omega) = \frac{1}{2K_a(\omega)}\langle \frac{1}{2} \{ X_{\phi_L}^{out},X_{\phi_L}^{out} \} \rangle= \frac{h^2_{SQL}}{2}   \left[ \frac{(K_a(\omega) - \cot \phi_L)^2 + 1}{K_a(\omega)} \right] $, where 
   $X_{\phi_L}^{out} = \frac{1}{\sqrt{2}} (a^{out} e^{-i\phi_L}+ {a^\dagger}^{out} e^{i\phi_L})$ is the output quadrature expressed in terms of output annihilation (creation) field operators: $a^{out}({a^\dagger}^{out})$. $K_a(\omega)$ is the optomechanical coupling strength and $\phi_L$ is the angle of output quadrature and $h_{SQL}$ is the standard quantum limit (SQL) of a free mass for GW strain \cite{Danilishin_GW_2019}}. This scheme is currently being installed in LIGO-Virgo. However, this approach is limited by the shot noise. Also, the response quadrature is extremely narrow and highly dependent on the operational frequency \cite{Danilishin_GW_2019}. The shot noise can be reduced by brutally increasing the power of the pump laser, but, that increases the QBA noise, as two noises are correlated. The QBA noise becomes significant when the measurement time goes long, i.e. it dominates at lower frequencies (below $\approx$ 100 Hz), as QBA noise spectral density is proportional to the susceptibility of the massive suspended mirror ($\chi_M = - 1/\omega^2$, where $\omega$ is the reference frequency in homodyne detection). In addition to above two dominant quantum noises the low-frequency band is also affected by seismic, electronic, and thermal noises.
   
   A further improvement in increasing the sensitivity is done by using squeezed light as a probe. { Using this idea, another scheme has been in a proposal for noise reduction, by injecting squeezed vacuum as an input ($ S_a (\omega) = \frac{h^2_{SQL}}{2}   \left[ e^{-2r}  K_a(\omega) + \frac{e^{2r}}{K_a(\omega)} \right] $, where $r$ is the squeezing parameter) \cite{LIGO_sensitivity_squeezing, Danilishin_GW_2019}.} Even though the input quadrature is frequency-dependent, the scheme has been able to reduce either the shot or QBA noise, but at the cost of an equivalent increment of each other. Therefore, the major problem still remains the reduction of both noises, simultaneously. Working in this direction, recently a highly interesting hybrid quantum scheme has been proposed, introducing a concept of the negative mass,  which involves an auxiliary system that consists of an atomic spin ensemble \cite{study4roadmap, Polzik_GW_Spin_PRD}. Here, a two-mode squeezed vacuum (TMSV) is sent to the spin-optomechanical hybrid system as an input, and the output response of the hybrid system is measured through heterodyne detection. This combined measurement suppresses both the shot noise and QBA noise by a factor $\cosh 2r$ (r is the squeezing amplitude), by back-action evading (BAE) measurements. Even though that this scheme has triggered high interests in the design of GW detectors, it faces severe implementation issues. The major issue is to design such a negative mass atomic spin system working at low frequencies and bandwidth so that both the susceptibilities of the spin system and mechanical mirror can match perfectly each other. The problem becomes more lethal, especially, at lower frequencies as the susceptibilities are more likely to deviate from each other \cite{study4roadmap, Polzik_GW_Spin_PRD}.
	
\begin{figure*}[t!]
		\includegraphics[width=12 cm]{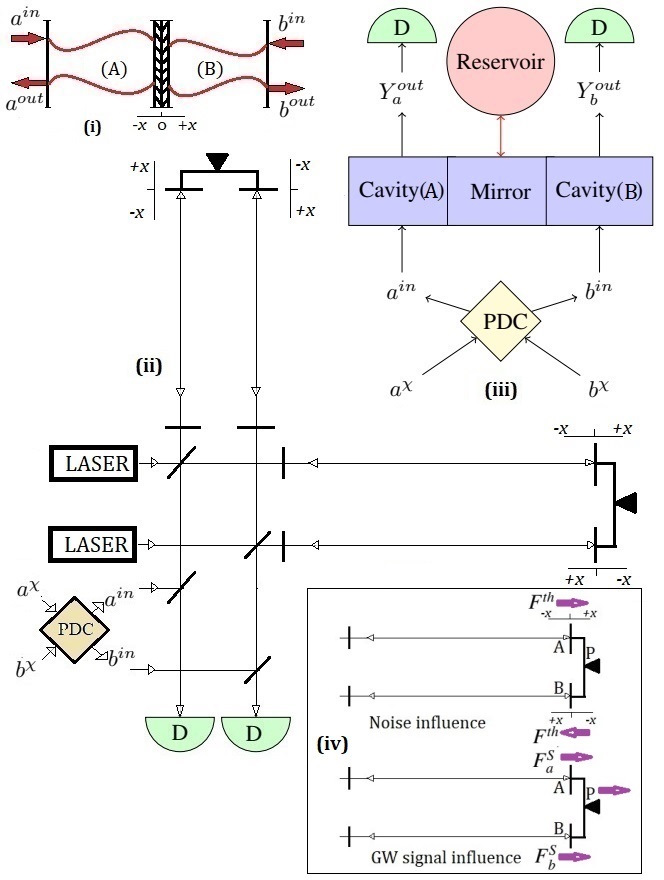}
		\caption{\label{block_GW_diagram} (i) Schematic diagram of the principle of the negative radiation pressure realized by sharing a common two-faced mechanical mirror. However, for practical realization, we replace this configuration with a scheme (ii) composed of two mirrors supported by a joint pivot. (ii) Implementation of the optomechanical double cavity, on a realistic GW interferometer; where the end mirrors are attached and suspended on a pivot so that they can move together in opposite directions. (iii) Block diagram of the double cavity BAE measurement setup, fed by input TMSV generated through the PDC process, and the outputs are detected through spatially separated heterodyne detection. {(iv) The top panel image shows the movement of the end mirrors on a pivot P for thermal noise and quantum back action noise and makes the end mirrors move in opposite directions cancelling this noise.  However, at the same time, the pivot P itself moves with respect to the front mirrors under the influence of GW tidal force with both end mirrors  in the same direction in the bottom (lower panel) .} }
	\end{figure*}

	To overcome these problems, here, we propose a new idea of introducing two cavities with mutually interlocked mirrors moving in opposite directions. We show, that by this design, which we call a negative radiation pressure coupling, one can very effectively suppress both the shot and QBA noise, simultaneously. This approach is based on geometrical considerations and it omits the auxiliary spin system \cite{study4roadmap}. Similarly to the spin-cavity hybrid system, the proposed scheme thus suppresses both noises simultaneously but with much higher efficiency, which we discuss below.  
{The scheme is significantly simpler to implement practically, by modifying the currently well-developed mirror hanging and suspension systems, such as in \cite{CQG_recent_2023}. The proposed principle is depicted in Fig. \ref{block_GW_diagram}(i). According to this scheme two photon flux are sent to two cavities which bounces back from same end mirror. The  end mirror is constructed by such a way that it is double faced so each part of the split beam arrives to the opposite face of the mirror. By this way the back action noise is compensated and we called this principle the negative radiation pressure coupling. However as this geometry is practically difficult to realize GW signal, instead, we work out further this principle by involving two cavities shown in Fig. \ref{block_GW_diagram}(ii). In this case, a second cavity is introduced by bringing another interferometer alongside the primary (main) one and attaching and suspending the end mirrors of a both cavities on a joint pivot, so that both mirrors can move in opposite directions with respect to each other under the influence of QBA and in the same direction ( due to the pivot movement) under the influence of GW. Our scheme proves to be significantly more efficient at low frequencies, especially when it comes to suppressing the QBA noise. Besides, our model is also capable of suppressing the thermal noise, which turns out to be an extra benefit of our scheme. The system works by injecting TMSV generated throughout parametric down conversions (PDC) as explained in detail below (Fig. \ref{block_GW_diagram}(iii)).\\	
    In some other approaches, local readout based upon a dual laser beam has been suggested in the context of the increment of the sensitivity of the GW detector  \cite{rev1_canceel_front_mirror}.  However, though this concept could remove the QBA noise from the front mirror, 
    the secondary laser beam does not resonate in the arm cavities, and so does not handle the noise produced by the end mirror. In another approach, a dual beam optical spring has been proposed to increase the sensitivity of the GW detector, through modifying mechanical susceptibility by tuning the power, detuning frequency and bandwidth which effectively acts as a negative inertia \cite{revc_negative_opt_spring}. However, this approach did not lead to suppressing QBA or the thermal noise but rather amplifying the signal. Another proposition, that has also been experimentally tested is based on injecting squeezed vacuum, frequency-dependent output squeezing \cite{revb_EPR_freq_sqz} but the suppression is the same as for the spin-optomechanical hybrid scheme \cite{study4roadmap} and effective in a comparatively narrower range of frequency. However, our work highly efficiently suppresses both quantum noises and in addition reduces thermal noise simultaneously. }

	In this letter, we start with a brief description of our double cavity mechanical system, followed by the method of transferring the input TMSV into the cavities and engineering the light-matter interaction to obtain squeezed output below SQL. The investigation ends up with a numerical analysis to see the scheme potential in the search for future implementations on advanced GW interferometer installations.

	\begin{figure}[t!]
		\includegraphics[width=9 cm]{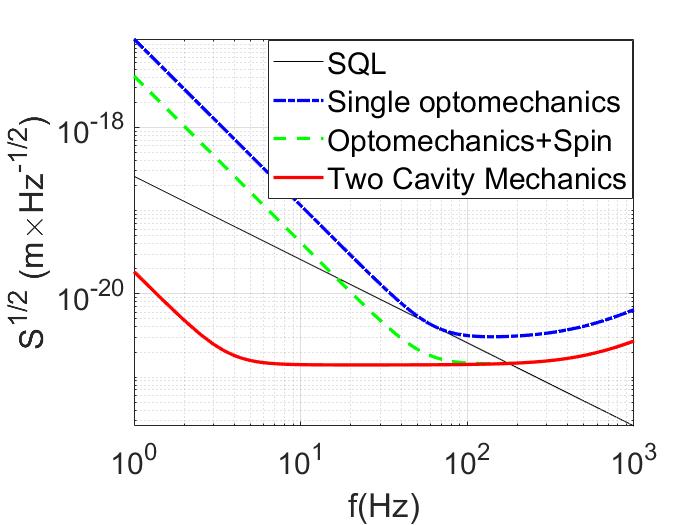}
		\caption{\label{output_sq_result} The spectral densities of quantum noise for $r= 1.7, L_{a} = 4000m, m= 40 kg, \kappa_{a}/2\pi = 500 Hz, \Theta_{a}/(2\pi)^3 = 100^3 s^{-3}$. In both cases, 2.5\% of input losses and
			2.5\% of output losses are assumed. In addition, for both optical cavities and spin system 0.01\% of intracavity roundtrip losses are assumed.}
	\end{figure}

	\section{Model} \label{Model}
	
	{
	As our dual cavity optomechanical system shares a common pivot (see Fig. \ref{block_GW_diagram}(i)), the increment of the length of one cavity created by the movement of the mirror decreases the length of the other. This brings opposite signs to the radiation pressure couplings between the mechanical oscillator and two optical cavities. However, under the influence of GW tidal force, both the cavities stretch or squeeze together simultaneously. Therefore, when it applies on realistic interferometer in Fig. \ref{block_GW_diagram}(ii),  
    under the tidal GW force, the pivot moves synchronously inducing the same length changes in both cavities (Fig. \ref{block_GW_diagram}(iv)). }
 
 The Hamiltonian of the entire system can be then described as:

	\begin{align}\label{optomechanical_Hamiltonian}
		\hat{H} &= \hbar \omega_a a^\dagger a  + \hbar \omega_b b^\dagger b+ \frac{p^2}{2m} +  \frac{m \omega^2_m x^2}{2} - \hbar \frac{\omega_a}{L_a}  x a^\dagger a  \nonumber \\
		+ & \hbar \frac{\omega_b}{L_b}  x b^\dagger b 	+ i\hbar \left( a^\dagger E_a e^{-i\omega_{L_a} t}  + b^\dagger E_b e^{-i\omega_{L_b} t} -h.c. \right)
	\end{align}
	
	where  $L_a$ and $L_B$ are the lengths, $\omega_a$ and $\omega_b$ are the frequencies, and $a(a^\dagger)$ and $b(b^\dagger)$ are the annihilation(creation) operators of the cavity modes A and B, respectively. $x$ and $p$ are position and momentum operators of the mechanical mode. The driving field with strength $(E_{a,b})$ and with frequencies $\omega_{L_a, L_b}$ is applied on optomechanical cavities, which in turn determines the normalized optical power $ \Theta_{a,b} =  \frac{8 \omega_{L_{a,b}} I_{a,b}}{mcL_{a,b}} $ in the linear coupling regime. $m, \omega_m$ are the mass and oscillation frequency of the mechanical resonator, respectively. The mechanical resonator and both the optical cavities interact continuously with their corresponding bosonic thermal bath which is of Markovian nature, and can be approximated as having a flat and infinite spectral density \cite{Cavity_optomechanics_Aspelmeyer}. 
	{Our scheme advances  a previous very interesting concept that was using an active feedback control to
		lock the motion of a mirror \cite{Courty_feedback_control, Vitali_feedback_control}. In this case, another photon beam impinging on the main interferometer mirror is fed through a small control mirror providing the feedback. The external feedback force is optimized to reduce optomechanical back action, which is controlled by the output response of cavity B. However, this concept is limited by the shot noise. Unlike the feedback control, our model, therefore, remains fundamentally different, as we used two cavities to generate the negative radiation pressure and TMSV as input to both cavities so that both the shot and QBA noise can be suppressed at the output. Our scheme has thus an important advantage of suppressing both the shot and QBA noise, simultaneously and to a degree significantly higher than another concept of negative mass \cite{study4roadmap}. In addition, our concept also reduces the thermal noise on mechanical mirrors at the output, which is a significant problem in the realistic construction of the GW interferometers. For example, approaches towards the reduction of thermal noise are being developed using low thermal noise coatings, but these are still rather far from practical realization. Here, with a rather simple optical concept, we can suppress these noises, in addition to back action noise and the shot noise.}
	
	Considering the interferometer tuned on resonance and without taking optical losses into account (see Appendix \ref{input_output_internal_losses} for details), each of the outputs of the interferometer are measured in their phase quadrature $(\phi_L = \pi/2)$ individually by setting up a homodyne detector, presented by the output quadratures as:

\begin{subequations}
\begin{align} \label{Y_outputs}
	Y_{a}^{out} &=  e^{2i \beta_{a}} Y_{a}^{in}  +  K_{ab} X_{b}^{in}  - K_{a} X_{a}^{in} 
	\\ &+  i \sqrt{2 K_{a} }  \frac{F^S_{a} + F^{th} }{F^{sql}  }   \, \, \text{for A},\nonumber \\
        Y_{b}^{out} & =  e^{2i \beta_{b}} Y_{b}^{in}  + K_{ab} X_{a}^{in}  - K_{b} X_{b}^{in} \\
	&+  i \sqrt{2 K_{b} }  \frac{F^S_{b} - F^{th} }{F^{sql}  } ,  \, \, \text{for B} \nonumber
\end{align}    
\end{subequations}

	where $	K_{a,b}(\omega) = - e^{(2i \beta_{a,b})} \frac{2 \kappa_{a,b} \Theta_{a,b} \chi_M}{\kappa_{a,b}^2 + \omega^2}$ and $K_{ab} = \sqrt{K_aK_b}$, $\kappa_{a,b}$ is the rate of dissipation of the corresponding cavity and $\beta_{a,b} = \arctan(\omega/\kappa_{a,b} )$ . $F^{th}$ is the thermal force acting on the mirror and $F^{S}_{a,b}$ is the signal force on each cavity which is invoked by the gravitational wave. {In the above equation, the GW force acts via the pivot movement in the same direction in both cavities, but the thermal as well as radiation pressure forces act in opposite directions.} $F^{sql} = \sqrt{\hbar m/\chi_M} $ is the standard quantum limit of the force which in turn fixes the minimum limit of the spectral density of the sum of the shot and QBA noises $(S^{sql}_F = \hbar m \omega^2 )$
	
	
	We probe two cavities of the optomechanical system with entangled TMSV as an input signal which is generated by non-linear optical crystal ($\chi$) through the PDC and measure the outputs parallel through heterodyne detection (Fig. \ref{block_GW_diagram}(iii)). Considering $r$ as the arbitrary amplitude of the squeezing parameter and fixing the phase to $\pi/2$ in this PDC process, {so that the quadratures $X^{in}_a$ and  $Y^{in}_a$ remain uncorrelated and their spectral densities become $e^{2r}/2$ and $e^{-2r}/2$}. The regenerated entangled modes satisfy

	\begin{align}\label{PDC_TMSV_quadrature}
		X^{in}_{a,b} &= \cosh r X^\chi_{a,b} + \sinh r X^\chi_{b,a} \nonumber \\
		Y^{in}_{a,b} &= \cosh r Y^\chi_{a,b} - \sinh r Y^\chi_{b,a},
	\end{align}
	
	where $ X^\chi_{a,b},  Y^\chi_{a,b}$ are the amplitude and phase quadratures of two independent vacuum modes, probed into the nonlinear crystal.

	{
	To suppress both the shot and QBA noises, we express the optomechanical coupling strengths to be $K_b = W K_a$, where $W$ is a weight factor. Hereafter, we parameterize the weight factor and minimize the quantum noise spectra with respect to it, which imposes a condition $K_b = (\tanh 2r)^2 K_a$, and therefore, $K_{ab} = (\tanh 2r) K_a$. Therefore, the condition mathematically provides the maximum suppression of the quantum noise. } This condition can be easily achieved by tuning the strength of the pump lasers. This leads to obtaining the combined output measurement, weighted by $\tanh 2r$ on the cavity $B$, as

	\begin{align} \label{Y_out_balanced}
		Y_{a}^{out} &+ Y_{b}^{out} \tanh 2r =  e^{2i \beta_{a}} Y_{a}^{in} +  e^{2i \beta_{b}} Y_{b}^{in} \tanh 2r  \nonumber \\ 
		&- \frac{ K_a }{\cosh^2 2r} ( X_a^{in} - \tanh 2r X_b^{in} )  \nonumber	\\
		&+ \frac{i\sqrt{2  K_a}}{\cosh^2 2r} \frac{F^{th}}{ F^{sql}} + i\sqrt{2  K_a} \frac{F^{S}_a+\tanh^2 2r F^{S}_b}{ F^{sql}},
	\end{align}
	
	which corresponds to the suppression of both shot and QBA noises by the factors $\cosh 2r$ and $\cosh^5 2r$, respectively {($ S (\omega) = \frac{1}{2K_a(\omega)}\langle \frac{1}{2} \{ Y_{a}^{out} + Y_{b}^{out} \tanh 2r,Y_{a}^{out} + Y_{b}^{out} \tanh 2r \} \rangle $)}. In addition, the first element of the third line of the Eq. \eqref{Y_out_balanced} shows that the scheme is also been able to suppress the thermal noise by $\cosh^4 2r$. 
{
Besides, as discussed before, the structure of the double cavity GW interferrometer in Fig.  \ref{block_GW_diagram}(ii) ensures both the end mirrors, including the suspension pivot, move same direction at a time under the influence of GW tidal force (Fig. \ref{block_GW_diagram}(iv)), which leads to satisfying $F^S_a = F^S_b$, for an equal length of the cavities A and B. This causes an enhancement in the balanced detection of GW signal with a factor $(1+\tanh^2 2r)^2$.
	}
 Note that optical losses are not taken into account so far in this theoretical model, which can easily be introduced as it is shown in \cite{study4roadmap}. The modifications of input/output formalism in the presence of losses, and the associated balancing conditions to obtain BAE measurements, are given in Appendix \ref{input_output_internal_losses}.	 While evaluating the efficiency of this scheme through numerical estimation in the next section, we included several possible kinds of optical losses corresponding to a realistic scenario.
	
	In order to implement the scheme in realistic cavity optomechanical systems, one can make the cavities as close to identical to each other, by engineering their finesse, cavity resonance frequencies, and the single photon coupling to the mechanical motion. As both cavities are stabilized by driving pump of the same frequencies ($\omega_{L_a} = \omega_{L_b}$), the deviation of the cavity parameters from their identical behavior can be adjusted by properly tuning the intensities of the pump lasers so that the balancing condition $K_b = (\tanh 2r)^2 K_a$ remains uncompromised.

	\section{Numerical Estimation}
	
	Using numerical simulations, we determined both the shot and QBA noise together and compare them to the well-balanced hybrid spin-optomechanical system, proposed in \cite{study4roadmap, Polzik_GW_Spin_PRD}, to visualize the efficiency of the scheme. In this simulation, as explained above, we consider the cavity B to be identical to the cavity A. Keeping in mind the possibility of a practical implementation, we have chosen modeling parameters corresponding to those of the LIGO interferometer and a realistic GW detector \cite{study4roadmap}.  As anticipated, Fig \ref{output_sq_result} shows that the shot and the QBA noise together are limited by the SQL for a simple cavity optomechanical system. However, in the case of the balanced hybrid spin-optomechanical system (for a situation when the susceptibility of atomic spin and mechanical mirror are perfectly matched), for input squeezing $r= 1.7$ ($\approx 15$ db) the fluctuation goes below SQL and the sensitivity gain is around 7.5 db across the entire frequency range of interest. As seen in Fig. 2, our double cavity optomechanical system scheme, reducing both noises, can suppress the fluctuation below SQL for a much larger range of frequencies. As anticipated, the reduction of the shot noise remains the same with the spin-optomechanical hybrid system ($\approx 7.5 db$), however the QBA noise reduction increases for our scheme to 56 db (9 db for the hybrid spin-optomechanical system). As the QBA noise dominates at a lower frequency, the scheme is seen to be much more efficient, especially at lower frequencies ( $<$ 100 Hz). 
	
	\section{Conclusion}
	
	We present a novel scheme suppressing at the same time both the shot and the optomechanical back action noise, reaching beyond the standard quantum limit. Compared to the schemes proposed before, our scheme appears to be much more efficient for, firstly, not demanding any auxiliary spin system, therefore there is no need to design a negative mass spin system with a lower Larmor frequency and bandwidth. Secondly, it has the ability to suppress the QBA noise {(56 db)} more than the spin-optomechanical hybrid scheme {(9db)}, along with the same rate of suppression of shot noise. Thirdly, it suppresses the thermal noise simultaneously with good efficiency. We believe that our scheme can bring remarkable advantages for designing the second generation GW detectors, operating especially at lower frequencies. 
	
	\begin{acknowledgments}
		SA would like to thank Philippe Djorwe for his fruitful suggestions. The work has been supported by the E-Test (EMR) project and European Union, Nuclear singlet state in diamond for overcoming the
		standard quantum limit in gravitational wave detectors (SingletSQL) GA no 101065991.
		
	\end{acknowledgments}

	\nocite{*}

	\bibliography{apssamp}

\begin{thebibliography}{21}%
\makeatletter
\providecommand \@ifxundefined [1]{%
 \@ifx{#1\undefined}
}%
\providecommand \@ifnum [1]{%
 \ifnum #1\expandafter \@firstoftwo
 \else \expandafter \@secondoftwo
 \fi
}%
\providecommand \@ifx [1]{%
 \ifx #1\expandafter \@firstoftwo
 \else \expandafter \@secondoftwo
 \fi
}%
\providecommand \natexlab [1]{#1}%
\providecommand \enquote  [1]{``#1''}%
\providecommand \bibnamefont  [1]{#1}%
\providecommand \bibfnamefont [1]{#1}%
\providecommand \citenamefont [1]{#1}%
\providecommand \href@noop [0]{\@secondoftwo}%
\providecommand \href [0]{\begingroup \@sanitize@url \@href}%
\providecommand \@href[1]{\@@startlink{#1}\@@href}%
\providecommand \@@href[1]{\endgroup#1\@@endlink}%
\providecommand \@sanitize@url [0]{\catcode `\\12\catcode `\$12\catcode
  `\&12\catcode `\#12\catcode `\^12\catcode `\_12\catcode `\%12\relax}%
\providecommand \@@startlink[1]{}%
\providecommand \@@endlink[0]{}%
\providecommand \url  [0]{\begingroup\@sanitize@url \@url }%
\providecommand \@url [1]{\endgroup\@href {#1}{\urlprefix }}%
\providecommand \urlprefix  [0]{URL }%
\providecommand \Eprint [0]{\href }%
\providecommand \doibase [0]{https://doi.org/}%
\providecommand \selectlanguage [0]{\@gobble}%
\providecommand \bibinfo  [0]{\@secondoftwo}%
\providecommand \bibfield  [0]{\@secondoftwo}%
\providecommand \translation [1]{[#1]}%
\providecommand \BibitemOpen [0]{}%
\providecommand \bibitemStop [0]{}%
\providecommand \bibitemNoStop [0]{.\EOS\space}%
\providecommand \EOS [0]{\spacefactor3000\relax}%
\providecommand \BibitemShut  [1]{\csname bibitem#1\endcsname}%
\let\auto@bib@innerbib\@empty
\bibitem [{\citenamefont {Khalili}\ and\ \citenamefont
  {Polzik}(2018)}]{study4roadmap}%
  \BibitemOpen
  \bibfield  {author} {\bibinfo {author} {\bibfnamefont {F.~Y.}\ \bibnamefont
  {Khalili}}\ and\ \bibinfo {author} {\bibfnamefont {E.~S.}\ \bibnamefont
  {Polzik}},\ }\href {https://doi.org/10.1103/PhysRevLett.121.031101}
  {\bibfield  {journal} {\bibinfo  {journal} {Phys. Rev. Lett.}\ }\textbf
  {\bibinfo {volume} {121}},\ \bibinfo {pages} {031101} (\bibinfo {year}
  {2018})}\BibitemShut {NoStop}%
\bibitem [{adv({\natexlab{a}})}]{advanced_LIGO}%
  \BibitemOpen
  \href@noop {} {\bibinfo {title} {https://www.ligo.caltech.edu}}
  ({\natexlab{a}})\BibitemShut {NoStop}%
\bibitem [{adv({\natexlab{b}})}]{advanced_Virgo}%
  \BibitemOpen
  \href@noop {} {\bibinfo {title} {https://www.virgo-gw.eu}}
  ({\natexlab{b}})\BibitemShut {NoStop}%
\bibitem [{adv({\natexlab{c}})}]{advanced_GEO}%
  \BibitemOpen
  \href@noop {} {\bibinfo {title} {http://www.geo600.org}}
  ({\natexlab{c}})\BibitemShut {NoStop}%
\bibitem [{\citenamefont {Menoni}(2022)}]{coating_thermal_noise}%
  \BibitemOpen
  \bibfield  {author} {\bibinfo {author} {\bibfnamefont {C.~S.}\ \bibnamefont
  {Menoni}},\ }in\ \href {https://doi.org/10.1364/LAOP.2022.Tu2A.1} {\emph
  {\bibinfo {booktitle} {Latin America Optics and Photonics (LAOP) Conference
  2022}}}\ (\bibinfo  {publisher} {Optica Publishing Group},\ \bibinfo {year}
  {2022})\ p.\ \bibinfo {pages} {Tu2A.1}\BibitemShut {NoStop}%
\bibitem [{\citenamefont {Danilishin}\ \emph {et~al.}(2019)\citenamefont
  {Danilishin}, \citenamefont {Khalili},\ and\ \citenamefont
  {Miao}}]{Danilishin_GW_2019}%
  \BibitemOpen
  \bibfield  {author} {\bibinfo {author} {\bibfnamefont {S.~L.}\ \bibnamefont
  {Danilishin}}, \bibinfo {author} {\bibfnamefont {F.~Y.}\ \bibnamefont
  {Khalili}},\ and\ \bibinfo {author} {\bibfnamefont {H.}~\bibnamefont
  {Miao}},\ }\href {https://doi.org/10.1007/s41114-019-0018-y} {\bibfield
  {journal} {\bibinfo  {journal} {Living Reviews in Relativity}\ }\textbf
  {\bibinfo {volume} {22}},\ \bibinfo {pages} {2} (\bibinfo {year}
  {2019})}\BibitemShut {NoStop}%
\bibitem [{\citenamefont {Collaboration}(2013)}]{LIGO_sensitivity_squeezing}%
  \BibitemOpen
  \bibfield  {author} {\bibinfo {author} {\bibfnamefont {T.~L.~S.}\
  \bibnamefont {Collaboration}},\ }\href
  {https://doi.org/10.1038/nphoton.2013.177} {\bibfield  {journal} {\bibinfo
  {journal} {Nature Photonics}\ }\textbf {\bibinfo {volume} {7}},\ \bibinfo
  {pages} {613} (\bibinfo {year} {2013})}\BibitemShut {NoStop}%
\bibitem [{\citenamefont {Zeuthen}\ \emph {et~al.}(2019)\citenamefont
  {Zeuthen}, \citenamefont {Polzik},\ and\ \citenamefont
  {Khalili}}]{Polzik_GW_Spin_PRD}%
  \BibitemOpen
  \bibfield  {author} {\bibinfo {author} {\bibfnamefont {E.}~\bibnamefont
  {Zeuthen}}, \bibinfo {author} {\bibfnamefont {E.~S.}\ \bibnamefont
  {Polzik}},\ and\ \bibinfo {author} {\bibfnamefont {F.~Y.}\ \bibnamefont
  {Khalili}},\ }\href {https://doi.org/10.1103/PhysRevD.100.062004} {\bibfield
  {journal} {\bibinfo  {journal} {Phys. Rev. D}\ }\textbf {\bibinfo {volume}
  {100}},\ \bibinfo {pages} {062004} (\bibinfo {year} {2019})}\BibitemShut
  {NoStop}%
\bibitem [{\citenamefont {Sider}\ \emph {et~al.}(2023)\citenamefont {Sider},
  \citenamefont {Fronzo}, \citenamefont {Amez-Droz}, \citenamefont {Amorosi},
  \citenamefont {Badaracco}, \citenamefont {Baer}, \citenamefont {Bertolini},
  \citenamefont {Bruno}, \citenamefont {Cebeci}, \citenamefont {Collette},
  \citenamefont {Ebert}, \citenamefont {Erben}, \citenamefont {Esteves},
  \citenamefont {Ferreira}, \citenamefont {Gatti}, \citenamefont {Giesberts},
  \citenamefont {Hebbeker}, \citenamefont {van Heijningen}, \citenamefont
  {Hennig}, \citenamefont {Hennig}, \citenamefont {Hild}, \citenamefont
  {Hoefer}, \citenamefont {Hoffmann}, \citenamefont {Jacques}, \citenamefont
  {Jamshidi}, \citenamefont {Joppe}, \citenamefont {Kuhlbusch}, \citenamefont
  {Lakkis}, \citenamefont {Lenaerts}, \citenamefont {Locquet}, \citenamefont
  {Loicq}, \citenamefont {Van}, \citenamefont {Loosen}, \citenamefont
  {Nesladek}, \citenamefont {Reiter}, \citenamefont {Stahl}, \citenamefont
  {Steinlechner}, \citenamefont {Steinlechner}, \citenamefont {Tavernier},
  \citenamefont {Teloi}, \citenamefont {Pérez},\ and\ \citenamefont
  {Zeoli}}]{CQG_recent_2023}%
  \BibitemOpen
  \bibfield  {author} {\bibinfo {author} {\bibfnamefont {A.}~\bibnamefont
  {Sider}}, \bibinfo {author} {\bibfnamefont {C.~D.}\ \bibnamefont {Fronzo}},
  \bibinfo {author} {\bibfnamefont {L.}~\bibnamefont {Amez-Droz}}, \bibinfo
  {author} {\bibfnamefont {A.}~\bibnamefont {Amorosi}}, \bibinfo {author}
  {\bibfnamefont {F.}~\bibnamefont {Badaracco}}, \bibinfo {author}
  {\bibfnamefont {P.}~\bibnamefont {Baer}}, \bibinfo {author} {\bibfnamefont
  {A.}~\bibnamefont {Bertolini}}, \bibinfo {author} {\bibfnamefont
  {G.}~\bibnamefont {Bruno}}, \bibinfo {author} {\bibfnamefont
  {P.}~\bibnamefont {Cebeci}}, \bibinfo {author} {\bibfnamefont
  {C.}~\bibnamefont {Collette}}, \bibinfo {author} {\bibfnamefont
  {J.}~\bibnamefont {Ebert}}, \bibinfo {author} {\bibfnamefont
  {B.}~\bibnamefont {Erben}}, \bibinfo {author} {\bibfnamefont
  {R.}~\bibnamefont {Esteves}}, \bibinfo {author} {\bibfnamefont
  {E.}~\bibnamefont {Ferreira}}, \bibinfo {author} {\bibfnamefont
  {A.}~\bibnamefont {Gatti}}, \bibinfo {author} {\bibfnamefont
  {M.}~\bibnamefont {Giesberts}}, \bibinfo {author} {\bibfnamefont
  {T.}~\bibnamefont {Hebbeker}}, \bibinfo {author} {\bibfnamefont {J.~V.}\
  \bibnamefont {van Heijningen}}, \bibinfo {author} {\bibfnamefont {J.-S.}\
  \bibnamefont {Hennig}}, \bibinfo {author} {\bibfnamefont {M.}~\bibnamefont
  {Hennig}}, \bibinfo {author} {\bibfnamefont {S.}~\bibnamefont {Hild}},
  \bibinfo {author} {\bibfnamefont {M.}~\bibnamefont {Hoefer}}, \bibinfo
  {author} {\bibfnamefont {H.-D.}\ \bibnamefont {Hoffmann}}, \bibinfo {author}
  {\bibfnamefont {L.}~\bibnamefont {Jacques}}, \bibinfo {author} {\bibfnamefont
  {R.}~\bibnamefont {Jamshidi}}, \bibinfo {author} {\bibfnamefont
  {R.}~\bibnamefont {Joppe}}, \bibinfo {author} {\bibfnamefont {T.-J.}\
  \bibnamefont {Kuhlbusch}}, \bibinfo {author} {\bibfnamefont {M.~H.}\
  \bibnamefont {Lakkis}}, \bibinfo {author} {\bibfnamefont {C.}~\bibnamefont
  {Lenaerts}}, \bibinfo {author} {\bibfnamefont {J.-P.}\ \bibnamefont
  {Locquet}}, \bibinfo {author} {\bibfnamefont {J.}~\bibnamefont {Loicq}},
  \bibinfo {author} {\bibfnamefont {B.~L.~L.}\ \bibnamefont {Van}}, \bibinfo
  {author} {\bibfnamefont {P.}~\bibnamefont {Loosen}}, \bibinfo {author}
  {\bibfnamefont {M.}~\bibnamefont {Nesladek}}, \bibinfo {author}
  {\bibfnamefont {M.}~\bibnamefont {Reiter}}, \bibinfo {author} {\bibfnamefont
  {A.}~\bibnamefont {Stahl}}, \bibinfo {author} {\bibfnamefont
  {J.}~\bibnamefont {Steinlechner}}, \bibinfo {author} {\bibfnamefont
  {S.}~\bibnamefont {Steinlechner}}, \bibinfo {author} {\bibfnamefont
  {F.}~\bibnamefont {Tavernier}}, \bibinfo {author} {\bibfnamefont
  {M.}~\bibnamefont {Teloi}}, \bibinfo {author} {\bibfnamefont {J.~V.}\
  \bibnamefont {Pérez}},\ and\ \bibinfo {author} {\bibfnamefont
  {M.}~\bibnamefont {Zeoli}},\ }\href
  {https://doi.org/10.1088/1361-6382/ace230} {\bibfield  {journal} {\bibinfo
  {journal} {Classical and Quantum Gravity}\ }\textbf {\bibinfo {volume}
  {40}},\ \bibinfo {pages} {165002} (\bibinfo {year} {2023})}\BibitemShut
  {NoStop}%
\bibitem [{\citenamefont {Rehbein}\ \emph {et~al.}(2007)\citenamefont
  {Rehbein}, \citenamefont {M\"uller-Ebhardt}, \citenamefont {Somiya},
  \citenamefont {Li}, \citenamefont {Schnabel}, \citenamefont {Danzmann},\ and\
  \citenamefont {Chen}}]{rev1_canceel_front_mirror}%
  \BibitemOpen
  \bibfield  {author} {\bibinfo {author} {\bibfnamefont {H.}~\bibnamefont
  {Rehbein}}, \bibinfo {author} {\bibfnamefont {H.}~\bibnamefont
  {M\"uller-Ebhardt}}, \bibinfo {author} {\bibfnamefont {K.}~\bibnamefont
  {Somiya}}, \bibinfo {author} {\bibfnamefont {C.}~\bibnamefont {Li}}, \bibinfo
  {author} {\bibfnamefont {R.}~\bibnamefont {Schnabel}}, \bibinfo {author}
  {\bibfnamefont {K.}~\bibnamefont {Danzmann}},\ and\ \bibinfo {author}
  {\bibfnamefont {Y.}~\bibnamefont {Chen}},\ }\href
  {https://doi.org/10.1103/PhysRevD.76.062002} {\bibfield  {journal} {\bibinfo
  {journal} {Phys. Rev. D}\ }\textbf {\bibinfo {volume} {76}},\ \bibinfo
  {pages} {062002} (\bibinfo {year} {2007})}\BibitemShut {NoStop}%
\bibitem [{\citenamefont {Khalili}\ \emph {et~al.}(2011)\citenamefont
  {Khalili}, \citenamefont {Danilishin}, \citenamefont {M\"uller-Ebhardt},
  \citenamefont {Miao}, \citenamefont {Chen},\ and\ \citenamefont
  {Zhao}}]{revc_negative_opt_spring}%
  \BibitemOpen
  \bibfield  {author} {\bibinfo {author} {\bibfnamefont {F.}~\bibnamefont
  {Khalili}}, \bibinfo {author} {\bibfnamefont {S.}~\bibnamefont {Danilishin}},
  \bibinfo {author} {\bibfnamefont {H.}~\bibnamefont {M\"uller-Ebhardt}},
  \bibinfo {author} {\bibfnamefont {H.}~\bibnamefont {Miao}}, \bibinfo {author}
  {\bibfnamefont {Y.}~\bibnamefont {Chen}},\ and\ \bibinfo {author}
  {\bibfnamefont {C.}~\bibnamefont {Zhao}},\ }\href
  {https://doi.org/10.1103/PhysRevD.83.062003} {\bibfield  {journal} {\bibinfo
  {journal} {Phys. Rev. D}\ }\textbf {\bibinfo {volume} {83}},\ \bibinfo
  {pages} {062003} (\bibinfo {year} {2011})}\BibitemShut {NoStop}%
\bibitem [{\citenamefont {Ma}\ \emph {et~al.}(2017)\citenamefont {Ma},
  \citenamefont {Miao}, \citenamefont {Pang}, \citenamefont {Evans},
  \citenamefont {Zhao}, \citenamefont {Harms}, \citenamefont {Schnabel},\ and\
  \citenamefont {Chen}}]{revb_EPR_freq_sqz}%
  \BibitemOpen
  \bibfield  {author} {\bibinfo {author} {\bibfnamefont {Y.}~\bibnamefont
  {Ma}}, \bibinfo {author} {\bibfnamefont {H.}~\bibnamefont {Miao}}, \bibinfo
  {author} {\bibfnamefont {B.~H.}\ \bibnamefont {Pang}}, \bibinfo {author}
  {\bibfnamefont {M.}~\bibnamefont {Evans}}, \bibinfo {author} {\bibfnamefont
  {C.}~\bibnamefont {Zhao}}, \bibinfo {author} {\bibfnamefont {J.}~\bibnamefont
  {Harms}}, \bibinfo {author} {\bibfnamefont {R.}~\bibnamefont {Schnabel}},\
  and\ \bibinfo {author} {\bibfnamefont {Y.}~\bibnamefont {Chen}},\ }\href
  {https://doi.org/10.1038/nphys4118} {\bibfield  {journal} {\bibinfo
  {journal} {Nature Physics}\ }\textbf {\bibinfo {volume} {13}},\ \bibinfo
  {pages} {776} (\bibinfo {year} {2017})}\BibitemShut {NoStop}%
\bibitem [{\citenamefont {Aspelmeyer}\ \emph {et~al.}(2014)\citenamefont
  {Aspelmeyer}, \citenamefont {Kippenberg},\ and\ \citenamefont
  {Marquardt}}]{Cavity_optomechanics_Aspelmeyer}%
  \BibitemOpen
  \bibfield  {author} {\bibinfo {author} {\bibfnamefont {M.}~\bibnamefont
  {Aspelmeyer}}, \bibinfo {author} {\bibfnamefont {T.~J.}\ \bibnamefont
  {Kippenberg}},\ and\ \bibinfo {author} {\bibfnamefont {F.}~\bibnamefont
  {Marquardt}},\ }\href {https://doi.org/10.1103/RevModPhys.86.1391} {\bibfield
   {journal} {\bibinfo  {journal} {Rev. Mod. Phys.}\ }\textbf {\bibinfo
  {volume} {86}},\ \bibinfo {pages} {1391} (\bibinfo {year}
  {2014})}\BibitemShut {NoStop}%
\bibitem [{\citenamefont {Courty}\ \emph {et~al.}(2003)\citenamefont {Courty},
  \citenamefont {Heidmann},\ and\ \citenamefont
  {Pinard}}]{Courty_feedback_control}%
  \BibitemOpen
  \bibfield  {author} {\bibinfo {author} {\bibfnamefont {J.-M.}\ \bibnamefont
  {Courty}}, \bibinfo {author} {\bibfnamefont {A.}~\bibnamefont {Heidmann}},\
  and\ \bibinfo {author} {\bibfnamefont {M.}~\bibnamefont {Pinard}},\ }\href
  {https://doi.org/10.1103/PhysRevLett.90.083601} {\bibfield  {journal}
  {\bibinfo  {journal} {Phys. Rev. Lett.}\ }\textbf {\bibinfo {volume} {90}},\
  \bibinfo {pages} {083601} (\bibinfo {year} {2003})}\BibitemShut {NoStop}%
\bibitem [{\citenamefont {Vitali}\ \emph {et~al.}(2004)\citenamefont {Vitali},
  \citenamefont {Punturo}, \citenamefont {Mancini}, \citenamefont {Amico},\
  and\ \citenamefont {Tombesi}}]{Vitali_feedback_control}%
  \BibitemOpen
  \bibfield  {author} {\bibinfo {author} {\bibfnamefont {D.}~\bibnamefont
  {Vitali}}, \bibinfo {author} {\bibfnamefont {M.}~\bibnamefont {Punturo}},
  \bibinfo {author} {\bibfnamefont {S.}~\bibnamefont {Mancini}}, \bibinfo
  {author} {\bibfnamefont {P.}~\bibnamefont {Amico}},\ and\ \bibinfo {author}
  {\bibfnamefont {P.}~\bibnamefont {Tombesi}},\ }\href
  {https://doi.org/10.1088/1464-4266/6/8/010} {\bibfield  {journal} {\bibinfo
  {journal} {Journal of Optics B: Quantum and Semiclassical Optics}\ }\textbf
  {\bibinfo {volume} {6}},\ \bibinfo {pages} {S691} (\bibinfo {year}
  {2004})}\BibitemShut {NoStop}%
\bibitem [{\citenamefont {Danilishin}\ and\ \citenamefont
  {Khalili}(2012)}]{Danilishin_GW_2012}%
  \BibitemOpen
  \bibfield  {author} {\bibinfo {author} {\bibfnamefont {S.~L.}\ \bibnamefont
  {Danilishin}}\ and\ \bibinfo {author} {\bibfnamefont {F.~Y.}\ \bibnamefont
  {Khalili}},\ }\href {https://doi.org/10.12942/lrr-2012-5} {\bibfield
  {journal} {\bibinfo  {journal} {Living Reviews in Relativity}\ }\textbf
  {\bibinfo {volume} {15}},\ \bibinfo {pages} {5} (\bibinfo {year}
  {2012})}\BibitemShut {NoStop}%
\bibitem [{\citenamefont {M{\o}ller}\ \emph {et~al.}(2017)\citenamefont
  {M{\o}ller}, \citenamefont {Thomas}, \citenamefont {Vasilakis}, \citenamefont
  {Zeuthen}, \citenamefont {Tsaturyan}, \citenamefont {Balabas}, \citenamefont
  {Jensen}, \citenamefont {Schliesser}, \citenamefont {Hammerer},\ and\
  \citenamefont {Polzik}}]{study4calculation}%
  \BibitemOpen
  \bibfield  {author} {\bibinfo {author} {\bibfnamefont {C.~B.}\ \bibnamefont
  {M{\o}ller}}, \bibinfo {author} {\bibfnamefont {R.~A.}\ \bibnamefont
  {Thomas}}, \bibinfo {author} {\bibfnamefont {G.}~\bibnamefont {Vasilakis}},
  \bibinfo {author} {\bibfnamefont {E.}~\bibnamefont {Zeuthen}}, \bibinfo
  {author} {\bibfnamefont {Y.}~\bibnamefont {Tsaturyan}}, \bibinfo {author}
  {\bibfnamefont {M.}~\bibnamefont {Balabas}}, \bibinfo {author} {\bibfnamefont
  {K.}~\bibnamefont {Jensen}}, \bibinfo {author} {\bibfnamefont
  {A.}~\bibnamefont {Schliesser}}, \bibinfo {author} {\bibfnamefont
  {K.}~\bibnamefont {Hammerer}},\ and\ \bibinfo {author} {\bibfnamefont
  {E.~S.}\ \bibnamefont {Polzik}},\ }\href
  {https://doi.org/10.1038/nature22980} {\bibfield  {journal} {\bibinfo
  {journal} {Nature}\ }\textbf {\bibinfo {volume} {547}},\ \bibinfo {pages}
  {191} (\bibinfo {year} {2017})}\BibitemShut {NoStop}%
\bibitem [{\citenamefont {Duan}\ \emph {et~al.}(2000)\citenamefont {Duan},
  \citenamefont {Giedke}, \citenamefont {Cirac},\ and\ \citenamefont
  {Zoller}}]{Duan_Inseparability_entanglement}%
  \BibitemOpen
  \bibfield  {author} {\bibinfo {author} {\bibfnamefont {L.-M.}\ \bibnamefont
  {Duan}}, \bibinfo {author} {\bibfnamefont {G.}~\bibnamefont {Giedke}},
  \bibinfo {author} {\bibfnamefont {J.~I.}\ \bibnamefont {Cirac}},\ and\
  \bibinfo {author} {\bibfnamefont {P.}~\bibnamefont {Zoller}},\ }\href
  {https://doi.org/10.1103/PhysRevLett.84.2722} {\bibfield  {journal} {\bibinfo
   {journal} {Phys. Rev. Lett.}\ }\textbf {\bibinfo {volume} {84}},\ \bibinfo
  {pages} {2722} (\bibinfo {year} {2000})}\BibitemShut {NoStop}%
\bibitem [{\citenamefont {Simon}(2000)}]{Simon}%
  \BibitemOpen
  \bibfield  {author} {\bibinfo {author} {\bibfnamefont {R.}~\bibnamefont
  {Simon}},\ }\href {https://doi.org/10.1103/PhysRevLett.84.2726} {\bibfield
  {journal} {\bibinfo  {journal} {Phys. Rev. Lett.}\ }\textbf {\bibinfo
  {volume} {84}},\ \bibinfo {pages} {2726} (\bibinfo {year}
  {2000})}\BibitemShut {NoStop}%
\bibitem [{\citenamefont {Giovannetti}\ and\ \citenamefont
  {Vitali}(2001)}]{Brownian_stochastic_force_mechanics_vitali}%
  \BibitemOpen
  \bibfield  {author} {\bibinfo {author} {\bibfnamefont {V.}~\bibnamefont
  {Giovannetti}}\ and\ \bibinfo {author} {\bibfnamefont {D.}~\bibnamefont
  {Vitali}},\ }\href {https://doi.org/10.1103/PhysRevA.63.023812} {\bibfield
  {journal} {\bibinfo  {journal} {Phys. Rev. A}\ }\textbf {\bibinfo {volume}
  {63}},\ \bibinfo {pages} {023812} (\bibinfo {year} {2001})}\BibitemShut
  {NoStop}%
\bibitem [{\citenamefont {Knyazev}\ \emph {et~al.}(2018)\citenamefont
  {Knyazev}, \citenamefont {Danilishin}, \citenamefont {Hild},\ and\
  \citenamefont {Khalili}}]{Speedmeter_EPR}%
  \BibitemOpen
  \bibfield  {author} {\bibinfo {author} {\bibfnamefont {E.}~\bibnamefont
  {Knyazev}}, \bibinfo {author} {\bibfnamefont {S.}~\bibnamefont {Danilishin}},
  \bibinfo {author} {\bibfnamefont {S.}~\bibnamefont {Hild}},\ and\ \bibinfo
  {author} {\bibfnamefont {F.}~\bibnamefont {Khalili}},\ }\href
  {https://doi.org/https://doi.org/10.1016/j.physleta.2017.10.009} {\bibfield
  {journal} {\bibinfo  {journal} {Physics Letters A}\ }\textbf {\bibinfo
  {volume} {382}},\ \bibinfo {pages} {2219} (\bibinfo {year} {2018})},\
  \bibinfo {note} {special Issue in memory of Professor V.B.
  Braginsky}\BibitemShut {NoStop}%
\end{thebibliography}%
	\bibliographystyle{apsrev4-2}
	
	\appendix
	
	\section{Inclution of input, output and internal losses} \label{input_output_internal_losses}
	
	Taking into account optical losses, the Eq. \eqref{Y_outputs} can be modified as 
	
	\begin{widetext}
		\begin{align}
			Y_{a,b}^{out} =& \sqrt{\eta_{a,b}^{out}} \big[ R_{a,b}  Y_{a,b}^{inp}  + \sqrt{\eta_{a}^{int} \eta_{b}^{int}} K_{ab} X_{b,a}^{inp}  - \eta_{a,b}^{int} K_{a,b} X_{a,b}^{inp} + T_{a,b} {v_Y}_{a,b}^{int}  + \sqrt{\eta_{a,b}^{int}   (1-\eta_{b,a}^{int})}  K_{ab} {v_X}_{b,a}^{int} \nonumber
			\\ &- \sqrt{\eta_{a,b}^{int}(1-\eta_{a,b}^{int})} K_{a,b} {v_X}_{a,b}^{int}	+  i \sqrt{2 \eta_{a,b}^{int} K_{a,b} }  \frac{F^S_{a,b} \pm F^{th} }{F^{sql}} \big] + \sqrt{1- \eta_{a,b}^{out}} {v_Y}_{a,b}^{out}
		\end{align}
	\end{widetext}
	where
	$R_{a,b} = \frac{2 \eta_{a,b}^{int} \kappa_{a,b} }{\kappa_{a,b} - i\omega} - 1$ and $T_{a,b} = \frac{2  \kappa_{a,b} \sqrt{\eta_{a,b}^{int}(1-\eta_{a,b}^{int})} }{\kappa_{a,b} - i \omega}$
	and
	
	\begin{subequations}
		\begin{align}
			X_{a,b}^{inp} = \sqrt{\eta_{a,b}^{inp} } X_{a,b}^{in} + \sqrt{ (1- \eta_{a,b}^{inp}) } {v_X}_{a,b}^{in}\\
			Y_{a,b}^{inp} = \sqrt{\eta_{a,b}^{inp} } Y_{a,b}^{in} + \sqrt{ (1- \eta_{a,b}^{inp}) } {v_Y}_{a,b}^{in}
		\end{align}
	\end{subequations}
	
	are the effective incident light quadratures for the interferometer cavities, ${v_X}_{a,b}^{in}$
	and ${v_Y}_{a,b}^{in}$, ${v_X}_{a,b}^{out}$
	and ${v_Y}_{a,b}^{out}$, and  ${v_X}_{a,b}^{int}$
	and ${v_Y}_{a,b}^{int}$ are the vacuum amplitude and phase noise operators associated with the input, output, and internal optical losses of the interferometer cavities, respectively, and $\eta_{a,b}^{inp}, \eta_{a,b}^{out}$ and $\eta_{a,b}^{int}$ are their corresponding quantum efficiencies. Therefore, the cross-correlation spectral densities of the input signals are

	\begin{subequations}
		\begin{align}
			2S[X_{a}^{inp},X_{a}^{inp}] = 2S[ Y_{a}^{inp},Y_{a}^{inp}] = 1 + 2 \eta_{a}^{inp} \sinh^2 r 		\\
			2S[ X_{b}^{inp},X_{b}^{inp}] =  2S[Y_{b}^{inp},Y_{b}^{inp}] = 1 + 2 \eta_{b}^{inp} \sinh^2 r 
		\end{align}
		
		\begin{align}
			2 S[X_{a}^{inp},  X_{b}^{inp}] = \sqrt{\eta_{a}^{inp}\eta_{b}^{inp} } \sinh 2 r
			\\
			2 S[Y_{a}^{inp},Y_{b}^{inp} ] = -\sqrt{\eta_{a}^{inp}\eta_{b}^{inp} } \sinh 2 r
		\end{align}
		
		\begin{align}
			S[X_{a}^{inp},  Y_{a}^{inp}] =  S[X_{a}^{inp},  Y_{b}^{inp}] =  S[ Y_{a}^{inp},   X_{b}^{inp}] \nonumber\\=  S[  X_{b}^{inp}, Y_{b}^{inp}] = 0
		\end{align}
	\end{subequations}
	
	Considering the input signal is a highly squeezed vacuum, i.e. very high value of $r$ and therefore $e^r$ dominates in the Eq. \eqref{PDC_TMSV_quadrature}, the impact of the vacuum noise operators is negligible with respect to the signal. In order to suppress QBA noise, it requires a modified weight factor $\sqrt{\frac{\eta_{a}^{out} \eta_{a}^{inp}}{\eta_{b}^{out}\eta_{b}^{inp}} } \tanh 2r$ on the measurements of the output of cavity B, without putting any weight on cavity A. It also requires a new balancing condition
	
	\begin{equation}
		K_b = \frac{\eta_{a}^{int} \eta_{a}^{inp} }{\eta_{b}^{int} \eta_{b}^{inp} } \tanh^2 2r K_a, \hspace{2mm} K_{ab} = \sqrt{ \frac{\eta_{a}^{int} \eta_{a}^{inp} }{\eta_{b}^{int} \eta_{b}^{inp} } } \tanh 2r K_a
	\end{equation}
	
	which suppresses the QBA noise by the factor $\frac{\cosh^5 2r}{\eta_{a}^{inp}}$

\section{Thermal force on mirror, pivot and coating} \label{Thermal_force_mirror_pivot_coating}

\begin{figure}[t!]
		\includegraphics[width=9 cm]{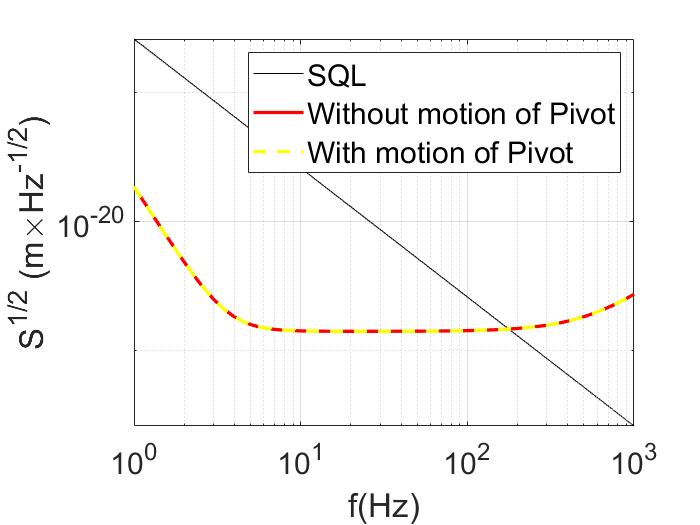}
		\caption{\label{pivot_motion} The spectral densities of quantum noise considering without and with the motion of pivot. The natural oscillation frequency and the frictional coefficient are considered as $ \omega_m = 0.25, \eta/m= 0.05$ Hz. The rest of all other parameters are considered the same with Fig. \ref{output_sq_result} }
	\end{figure}

{
  The thermal noise on the end mirror majorly has three different sources: the Brownian motion of the individual mirrors and the coatings on the mirrors and the thermal oscillations of the pivot. The thermal forces on individual mirrors which have already been taken into account in Eq. \eqref{Y_out_balanced}, are seen to be suppressed by the factor $\cosh^2 2r$. Here, we discuss the possibilities of two other sources of noise. The thermal force acting on the suspension pivot also contributes to the noise, which can be taken care of by considering the signal force on each cavity modified by $F^S_a = F^S_b = ( F^{GW} + F^{th}_{pivot})$ where $F^{GW}$ and $F^{th}_{pivot}$ are the forces by GW signal and thermal force on pivot, respectively. This modifies the Eq. \eqref{Y_out_balanced}, as
\begin{align} 
	Y_{a}^{out} &+ Y_{b}^{out} \tanh 2r =  e^{2i \beta_{a}} Y_{a}^{in} +  e^{2i \beta_{b}} Y_{b}^{in} \tanh 2r  \nonumber \\ 
	&- \frac{ K_a }{\cosh^2 2r} ( X_a^{in} - \tanh 2r X_b^{in} ) + \frac{i\sqrt{2  K_a}}{\cosh^2 2r} \frac{F^{th}}{ F^{sql}}  \nonumber	\\
	&+ i\sqrt{2  K_a} \frac{(1+\tanh^2 2r) ( F^{GW} +F^{th}_{pivot}) }{ F^{sql}}
\end{align} 
	which shows, even though the thermal forces acting on the individual mirror are being suppressed by the factor $\cosh^2 2r$, the thermal force on the pivot has been increased  by  $1+ \tanh^2 2r$. The thermal force on the pivot is dependent on temperature and frequency. Therefore, one way to decrease thermal noise is by increasing the stiffness of its suspension $(k \propto \omega^2$ and $n_{th} \propto  e^{-\beta \omega} )$. However, the higher operational frequency can affect the susceptibility $(\chi_M)$. In particular, the natural oscillation frequency and the damping of the suspension pivot has an influence on the susceptibility as $\chi_M = -\left(-\omega_m^2+\omega  (\omega +i \eta/m )\right)^{-1},$ where $\omega_m, m$ and $\eta $ are respectively the natural frequency, mass and the frictional coefficient of the mirror of the GW detector. In practical scenarios, detectors are designed where the mechanical mirror is hung through a multistage pendulum to obtain seismic isolation. For example, Virgo mirrors are suspended via a long chain of pendulum, hosted in towers 10 meters tall, therefore the natural frequency of the mirror is much lesser than the frequency of interest $(\omega_m << \omega)$. For example, in the case of the E-TEST project \cite{CQG_recent_2023},  0.23 Hz will be the frequency for the longitudinal and transversal modes. Heavy masses of the mirrors (40 kg for LIGO and 100 kg planned for Einstein telescope) will lead to suppressing the frictional losses ( by a high-quality factor). 
 In order to make the detector efficient at low frequencies ($\omega< 100$ Hz) the influence of susceptibility on the the noise should be minimal ($<<\omega$). Anticipating that the thermal noise ($n_{th} \propto  e^{-\beta \omega}$) will be dealt with by designing the natural frequency of the suspension and therefore the susceptibility, we calculate and plot the quantum noise with and without considering motions of the suspension pivot in Fig. }\ref{pivot_motion} . From this plot  we can deduce that the changes of quantum noise, considering the mechanical parameters of the mirrors and suspension appear to be minimal and can be easily be neglected. Further on, the thermal noise can be additionally reduced by controlling temperature, e.g. by a cryogenic cooler.

\end{document}